\documentclass[conference]{IEEEtran}
\IEEEoverridecommandlockouts
\usepackage{epsfig, endnotes}
\usepackage{amsmath}
\usepackage{amsthm}
\usepackage{amsmath,amsfonts}
\usepackage{algorithm}
\usepackage{algpseudocode}
\usepackage{array}
\usepackage{textcomp}
\usepackage{stfloats}
\usepackage{verbatim}
\usepackage{graphicx}
\usepackage{multirow}
\usepackage{graphicx} 
\usepackage{caption}  
\usepackage{booktabs}
\usepackage{xcolor}
\usepackage{soul}
\usepackage{tabularx}
\usepackage{lipsum}  
\usepackage{bm}
\usepackage{caption}
\usepackage{subcaption}
\usepackage{mathtools}
\usepackage{amsthm}

\newtheorem{theorem}{Theorem}

\newtheorem{assumption}{Assumption}  
\theoremstyle{remark}
\theoremstyle{claim}

\usepackage{url}

\renewcommand{\theequation}{\arabic{equation}}

\newcommand{\myparatight}[1]{\smallskip\noindent{\bf {#1}:}~}

\usepackage{xspace}

\usepackage{cleveref} 
\crefname{axiom}{axiom}{axioms} 
\crefname{definition}{definition}{definitions}
\crefname{lemma}{lemma}{lemmata}

\allowdisplaybreaks[3]

\begin{document}

\setlength{\textfloatsep}{4pt}
\setlength{\intextsep}{4pt}
\setlength{\floatsep}{4pt}

\title{On Transferring, Merging, and Splitting Task-Oriented Network Digital Twins
}

        

\author{
		\IEEEauthorblockN{Zifan Zhang,
        Minghong Fang,~\IEEEmembership{Member,~IEEE,} 
        Mingzhe Chen,~\IEEEmembership{Member,~IEEE,} 
        Yuchen Liu,~\IEEEmembership{Member,~IEEE} \\  
        }

  \thanks{Z. Zhang and Y. Liu are with the Department of Computer Science, North Carolina State University, Raleigh, NC, 27695, USA (Email: \{zzhang66, yuchen.liu\}@ncsu.edu). \textit{(Corresponding author: Yuchen Liu.)}}
  \thanks{M. Fang is with the Department of Computer Science and Engineering, University of Louisville, Louisville, KY, 40208, USA (Email: \{ minghong.fang@louisville.edu)\}.}
\thanks{M. Chen is with the Department of Electrical and Computer Engineering and Frost Institute for Data Science and Computing, University of Miami, Coral Gables, FL 33146 USA (Email: \protect\url{ mingzhe.chen@miami.edu)}.} 
}
\vspace{-1.5em}

\maketitle

\begin{abstract}
The integration of digital twinning technologies is driving next-generation networks toward new capabilities, allowing operators to thoroughly understand network conditions, efficiently analyze valuable radio data, and innovate applications through user-friendly, immersive interfaces. Building on this foundation, network digital twins (NDTs) accurately depict the operational processes and attributes of network infrastructures, facilitating predictive management through real-time analysis and measurement. However, constructing precise NDTs poses challenges, such as integrating diverse data sources, mapping necessary attributes from physical networks, and maintaining scalability for various downstream tasks. Unlike previous works that focused on the creation and mapping of NDTs from scratch, we explore intra- and inter-operations among NDTs within an Unified Twin Transformation (UTT) framework, which uncovers a new computing paradigm for efficient transfer, merging, and splitting of NDTs to create task-oriented twins. By leveraging joint multi-modal and distributed mapping mechanisms, UTT optimizes resource utilization and reduces the cost of creating NDTs, while ensuring twin model consistency. A theoretical analysis of the distributed mapping problem is conducted to establish convergence bounds for this multi-modal gated aggregation process. Evaluations on real-world twin-assisted applications, such as trajectory reconstruction, human localization, and sensory data generation, demonstrate the feasibility and effectiveness of interoperability among NDTs for corresponding task development.

\end{abstract}


\section{Introduction}

In the domain of telecommunications, wireless networks are experiencing a paradigmatic evolution, driven by the integration of advanced technologies such as edge computing~\cite{wang2023wireless}, millimeter-wave communication~\cite{10230133}, and machine learning~\cite{zhang2024REC}. These technologies are instrumental in laying groundwork for an array of novel applications and services in mixed physical and digital contexts, boosting capabilities of mobile broadband and enabling thorough integration of cyber-physical interactive systems.
Notably, the emergence of digital twins (DTs) marks a significant milestone in the digital transformation landscape, leveraging technologies such as Internet of Things (IoT), Artificial Intelligence (AI), and advanced analytics to create precise virtual replicas of physical systems or entities. These DTs facilitate real-time interactive correspondence between the physical world and their digital counterparts, enabling predictive analysis, emulation, and management through integrated data, models, and interfaces~\cite{digitaltwin5g, connectingtwins}. 
Building on this foundation, network digital twins (NDTs) extend the concept specifically to wireless networks~\cite{liu2024digital}, accurately depicting the operational processes and attributes of network infrastructures, including characteristics of physical and logical network devices and links, as defined in the International Telecommunication Union (ITU) standards~\cite{Y3090, Y3091}. Grounded in continuous mapping between the physical and digital realms~\cite{zhang2024mapping}, NDTs 
allow operators to thoroughly understand network's condition, analyze valuable radio data, and innovate network applications through user-friendly, immersive interfaces.

\myparatight{Motivation and Challenges} However, constructions of NDTs faces numerous challenges, primarily centered around the integration of diverse data sources, which is essential for accurately reflecting complex dynamics of wireless edges. This integration involves processing a wide spectrum of visual and sensory data, from system parameters to user behaviors and environmental factors, making it a \textit{data-intensive} task. Additionally, achieving real-time data synchronization between the physical network and its digital counterpart is imperative for maintaining an accurate model representation. Synchronization requires continuous communication mechanisms, posing significant data exchange and \textit{resource-related challenges}. 
This results in substantially high costs associated with processing multi-modal data, making it challenging to balance trade-offs between NDT model accuracy and creation efficiency. As demonstrated in prior works~\cite{liu2024digital, ramu2022federated, sun2020dynamic}, end-to-end twinning process from scratch incurs significant computational and communication overhead. Scalability also emerges as a significant concern, as NDTs must be constantly updated and adapted to remain accurate -- a \textit{computation-intensive} task given the complexity and scale of contemporary network operations.

\myparatight{Limitations and Opportunities} Due to these obstacles, developing a single universal NDT capable of performing \textit{every} task within a given domain is challenging under the variability and intricacy of network systems. A practical and efficient approach involves developing multiple functional, task-oriented NDTs, each focusing on specific tasks. These specialized twins, built using a consistent underlying mapping methodology, can be called upon as agents to perform their designated tasks when needed. Such a modular approach simplifies the design and maintenance of NDTs and enhances their flexibility and adaptability. Although several recent works have studied the creation and mapping of DTs from scratch, as noted in~\cite{zhang2024mapping, villa2024colosseum, 10255711}, a significant research gap remains in the \textit{intra- and inter-operations} among constructed DTs, 
which uncovers new paradigms for creating and managing NDTs, e.g. through joint data-model level transfer, merging, and splitting, to overcome the data, resource, and computation-intensive challenges.
On the other hand, addressing these interoperability issues is crucial for realizing the full potential of a networked system of DTs, thereby achieving coherent decision-making across different functional twins and optimizing the overall efficiency of network management.

\begin{figure}
	\centering
	\includegraphics[scale = 0.45]{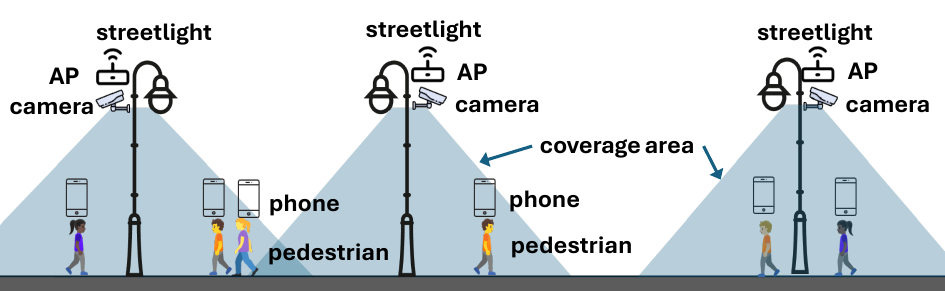}
	\caption{Example scenario for employing UTT framework with a distributed multi-modal mapping mechanism.}
	\label{fig: setting}
\end{figure}

\myparatight{Objectives and Contributions} 
To this end, this work aims to explore the feasibility of advanced interoperations, such as transferring, merging, and splitting multiple existing NDTs, to efficiently generate new task-oriented twins in a cost-efficient manner. 
This is achieved through a novel distributed multi-modal twinning framework called \textit{Unified Twin Transformation} (UTT), designed to effectively capture and leverage data and knowledge commonalities among twin models. 
In essence, two core techniques are developed in our UTT framework. First, a multi-modal mapping module enables integration of diverse data sources, resulting in a comprehensive and accurate representation of physical networks, analogous to how the human brain processes multi-modal information. Just as the brain can receive input in one modality, such as audio, and then generate outputs in other modalities like visual images and text, our multi-modal mapping scheme captures underlying associations between different data patterns within NDTs, which facilitates the analysis and understanding of complex tasks within same physical network system. This capability enables direct twin-to-twin transformation, which is significantly more cost-effective than conventional physical-to-twin creation.
Second, consider real-life network scenarios depicted in Fig.~\ref{fig: setting}. Each wireless access point (AP) or sensory device typically has limited coverage and can only manage data from a specific area. 
However, the overarching goal is to aggregate data from entire physical environment and enable collaborative operation among devices to construct task-oriented NDTs.
This naturally motivates the use of a federated learning framework for twinning operations, i.e., from several local-area task twins to a new global task twin. Through elaborate segmentation of network areas, 
our NDT transformation process leverages distributed learning from multiple local twins to capture geographically independent data and features, addressing heterogeneous multi-modal optimization problems in typical segmented wireless contexts. This implicitly decentralizes execution of NDT construction and downstream tasks across a network of devices. 
This workflow, akin to collaborative group brainstorming, employs a model aggregation mechanism to optimize the integration of diverse information pieces, which facilitates creation of new task-oriented twins from a global perspective.
Our main contributions can be summarized as follows:

\begin{list}{\labelitemi}{\leftmargin=1em \itemindent=-0.08em \itemsep=.2em}
    \item We introduce the first-of-its-kind UTT framework, which enables effective twin-to-twin operations such as transfer, merging, and splitting of existing NDTs. This facilitates the creation of new task-oriented DTs by sharing and transforming data knowledge. 
    
    \item We push forward the interoperability and accuracy of NDTs by leveraging a multi-modal mapping scheme. This allows for capturing underlying associations between different data patterns within NDTs constructed from diverse data sources.

    \item We adapt distributed learning to collaboratively tackle joint multi-modal, multi-area twinning problems, enhancing the robustness of the inter-twin transformation through a decentralized construction across a network of devices. 
    Theoretical analysis of this distributed mapping problem is also provided to establish convergence bounds for such a multi-modal gated aggregation.

    \item We conduct comprehensive evaluations on the performance of both twinning operations and multiple twin-assisted downstream tasks, such as human trajectory reconstruction, positioning, and inertial data generation, thereby validating the effectiveness of each task-specific NDT constructed.
\end{list}

\section{Preliminaries and Related Works}

\subsection{Network Digital Twins}

The critical role of NDTs in facilitating the next generation of networking technologies is emphasized in \cite{zhang2025synergizing, networkdigitaltwin, digitaltwin5g}, particularly in their impact on the strategic deployment of 5G networks to maximize coverage and data throughput. 
Despite most existing works primarily focusing on leveraging NDTs to optimize wireless networks, they often overlook details of creating DTs of physical infrastructure and properties~\cite{khan2022digital, zhang2024securing}. This twin construction process is a crucial prerequisite for subsequent applications of DTs in the field. 
Recently, a few works have delved into the study of constructing NDTs from scratch, involving data collection, differentiable modeling, and real-time refinement as in~\cite{villa2024colosseum, 10255711}. 
\cite{zhang2024mapping} also introduces a learning-based mapping strategy to construct NDTs with both historical data and newly coming real-time data. Our work explores a new cost-effective DT construction method through direct twin-to-twin transformation and interoperability. This approach does not conflict with conventional twinning methods but complements them, offering a cost-performance solution when several DTs are available and exploiting data and task-level commonalities.

\subsection{Multi-modal Intelligence for Integration}
Multi-modal learning has become a powerful approach in artificial intelligence, leveraging diverse data types such as text, images, and audio to improve the performance and reliability of traditional intelligent systems~\cite{huang2021makes,wu2025digital}. 
By leveraging the complementary information from different data sources, multi-modal learning can capture a more comprehensive representation of the underlying physical network and provide deeper insights into the system's behavior. 
This holistic understanding enhances the decision-making processes and operational strategies derived from the NDTs.
At the data level, \cite{xu2023multimodal} highlights transformers for multi-modal learning, while \cite{hong2020more} demonstrates the advantages of integrating diverse data sources, such as in remote-sensing imagery classification, showing that data diversity improves outcomes.
At the task level, \cite{yu2021learning} explores the capabilities of multi-modal learning to generate modality-specific representations through self-supervised multi-task learning. This technique not only refines the processing of each modality but also boosts the system's performance in tasks like sentiment analysis by utilizing the complementary strengths of the combined data types.
These studies motivate the adaptation of multi-modal intelligence into the digital twinning process, given its ability to manage complex and diverse datasets. Technically, the following three main stages can not only enhance twinning efficiency but also facilitate twin model integration for versatile downstream tasks across a wide array of disciplines.

\begin{list}{\labelitemi}{\leftmargin=1em \itemindent=-0.08em \itemsep=.2em}
    \item \textbf{Stage I (Data Representation)}: Each modality \( m \in M \) is represented in the data space using a feature extraction function \( \phi_m \), which transforms the raw collected data in \( m \) from a twin into a feature vector \( \mathbf{f}_m \).

    \item \textbf{Stage II (Modality Fusion)}: The feature vectors from different task-oriented twins are combined into a unified representation. A common fusion method is 
    $\mathbf{f}_{\text{fusion}} = [\mathbf{f}_1; \mathbf{f}_2; \ldots; \mathbf{f}_n],$
    where \( [\cdot; \cdot] \) denotes vector concatenation.

    \item \textbf{Stage III (Modal Integration)}: The fused feature vector \( \mathbf{f}_{\text{fusion}} \) is then used to train a knowledge model \( \bm{\theta} \), which can learn to perform specific tasks based on the integrated information from all data modalities and task behaviors.

\end{list}

\begin{figure*}[ht]  
    \centering
    \begin{subfigure}[b]{0.48\textwidth}  
        \centering
        \includegraphics[scale=0.4]{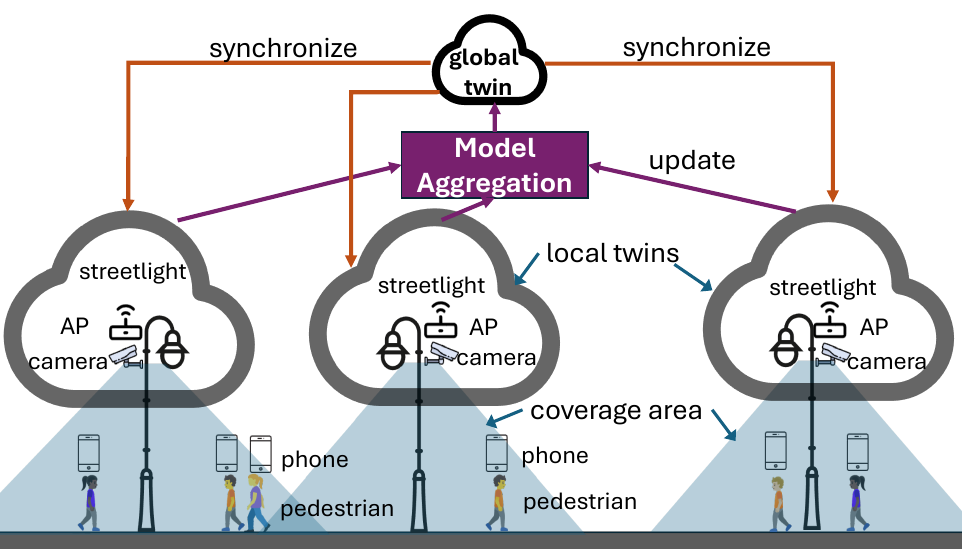}
        \caption{Architecture of UTT framework from a distributed transformation perspective.}
        \label{fig:fed}
    \end{subfigure}
    \hfill  
    \begin{subfigure}[b]{0.48\textwidth}  
        \centering
        \includegraphics[scale=0.4]{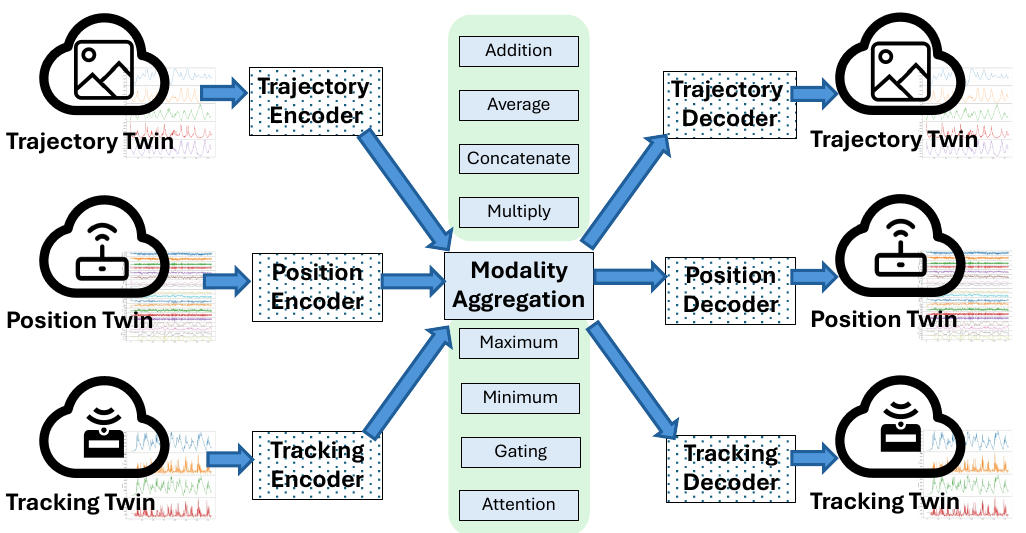}
        \caption{Architecture of UTT framework from a multi-modal transformation perspective.}
        \label{fig:modal}
    \end{subfigure}
    \vspace{-0.1in}
    \caption{The Overall architecture of UTT framework.}
    \vspace{-0.2in}
\end{figure*}

\vspace{-0.3cm}
\subsection{Distributed Learning for Model Adaptation}

Distributed learning allows for the parallel processing of large amounts of data collected from various network edges or areas, enabling real-time adaptation to the NDT by continuously learning from the latest data collected across APs and sensors, as highlighted in the ITU standards~\cite{Y3090,Y3091}.
Many studies, such as \cite{niknam2020federated, zhao2021federated, chen2020convergence}, adopt distributed learning to handle various datasets efficiently while significantly reducing bandwidth usage, shedding light on challenges such as data heterogeneity and communication overhead.
Given these prior studies, we previously developed vertical and horizontal mapping methods for constructing NDTs with a distributed learning scheme~\cite{zhang2024mapping}. This work explores feasibility of using federated mapping framework for \textit{interoperations} among NDTs, such as combining several local-area task twins into a new global task twin.
This can be formulated as an optimization problem, aiming to find new twin model parameters $\bm{\theta}^*$.
The optimization process unfolds in a distributed manner, jointly aggregating multi-modal ($n$) and multi-area ($z$) data, through a series of coordinated steps across each training round $t$:
\begin{list}{\labelitemi}{\leftmargin=1em \itemindent=-0.08em \itemsep=.2em}
    \item \textbf{Stage I (Synchronization):} The central mapping server distributes latest global task model $\bm{\theta}^t$ to each participating local-area twins.
    \item \textbf{Stage II (Local Model Training):} Each twin $i$ uses its local data in conjunction with current global twin model to update its local model, and then sends this updated model $\bm{\theta}_i^{t}$ back to the central server.
    \item \textbf{Stage III (Model Aggregation):} Using an aggregation rule, the central server combines all received local twin models to update the global twin model, e.g., to average all model parameters of local-area twins on each dimension to create a new twin. 
\end{list}

\begin{figure}[ht]
    \centering
    \vspace{-0.1in}
    \begin{subfigure}[b]{0.40\textwidth}
        \centering
        \includegraphics[width=\textwidth]{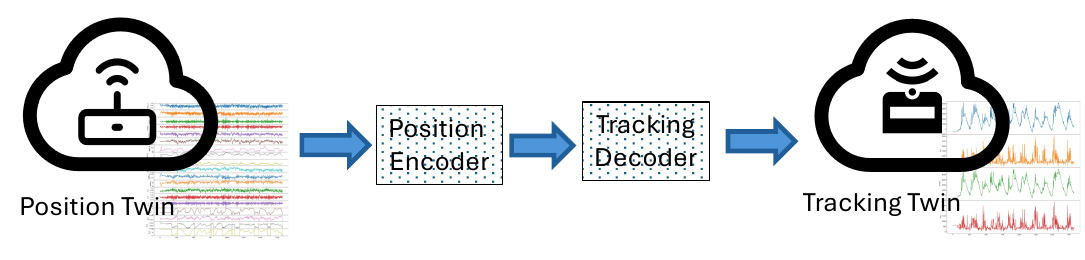}
        \caption{Transfer from Position Twin to Tracking Twin.}
        \label{fig:transfer1}
    \end{subfigure}
    \begin{subfigure}[b]{0.40\textwidth}
        \centering
        \includegraphics[width=\textwidth]{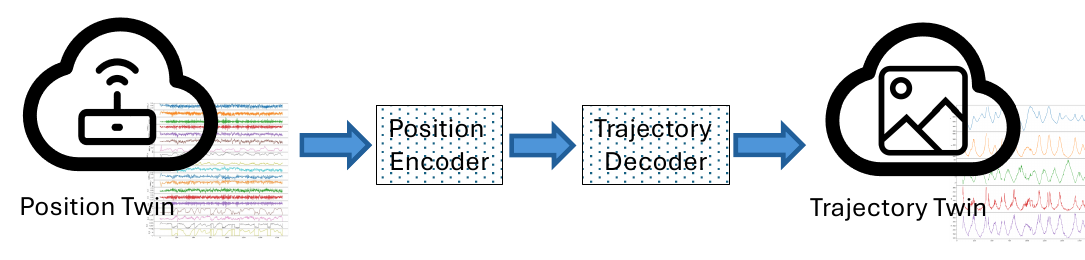}
        \caption{Transfer from Position Twin to Trajectory Twin.}
        \label{fig:transfer2}
    \end{subfigure}


    \begin{subfigure}[b]{0.23\textwidth}
        \centering
        \includegraphics[width=\textwidth]{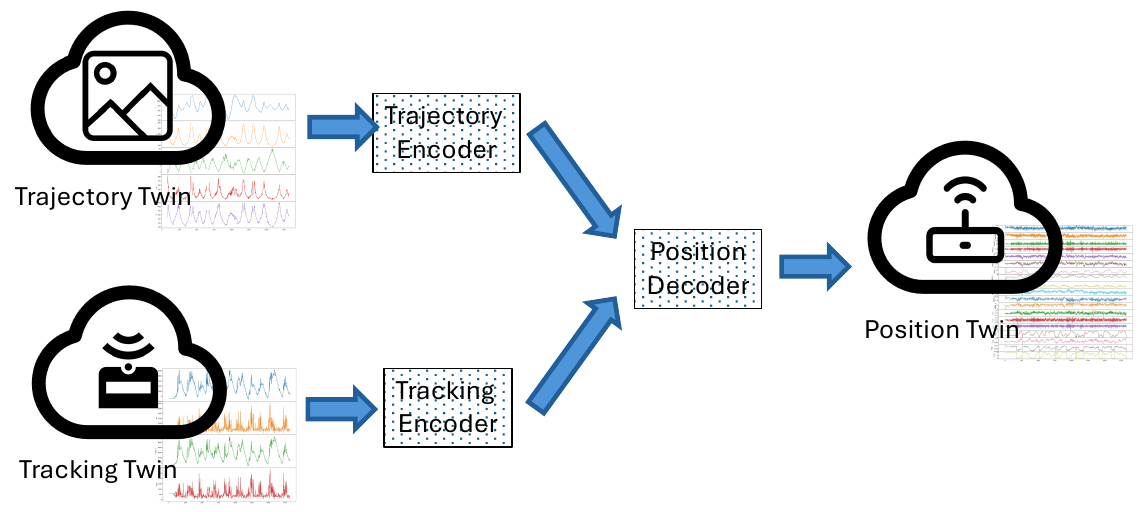}
        \caption{Merge Trajectory and Tracking Twins into Position Twin.}
        \label{fig:merge}
    \end{subfigure}
    \begin{subfigure}[b]{0.23\textwidth}
        \centering
        \includegraphics[width=\textwidth]{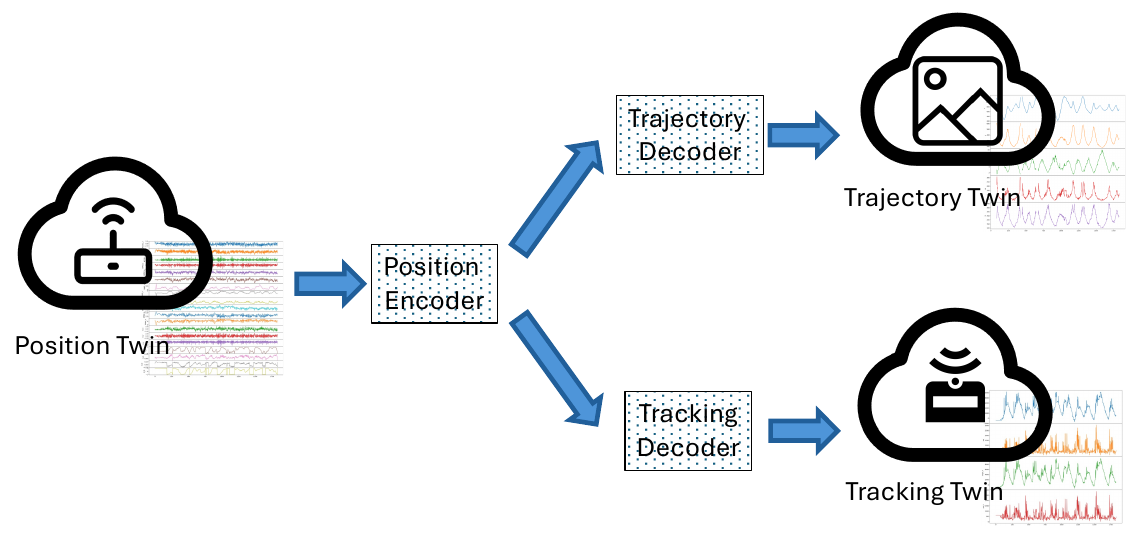}
        \caption{Split Position Twin into Trajectory and Tracking Twins.}
        \label{fig:split}
    \end{subfigure}
    \vspace{-.1in}
    \caption{Twin transformation functions enabled by UTT framework (using $\bm{\theta}_{Pos}^t$, $\bm{\theta}_{Pos}^t$, and $\bm{\theta}_{Traj}^t$ as an example).}
    \label{fig:multi_modal_learning}
    \vspace{-.1in}
\end{figure}
\vspace{-.1in}
\section{Unified Twin Transformation (UTT)}

Building upon the above preliminaries and analysis, we propose a novel distributed multi-modal twinning framework to enable effective intra- and inter-operations of constructed NDTs. This framework facilitates the creation of new task-oriented NDTs by capturing data knowledge commonalities from other NDTs within similar settings. 
Transferring knowledge from existing NDTs within the same environment preserves critical system dynamics, operational parameters, and historical data. This accumulated insight ensures the new NDT aligns with network-wide characteristics, reducing inconsistencies and representation gaps.

The realization of such NDT transformations presents two significant technical challenges. First, finding a unified representation for complex wireless networks is non-trivial. Integrating multiple data modalities—ranging from high-dimensional system parameters to diverse environmental and behavioral signals—requires sophisticated mapping algorithms capable of reconciling heterogeneous data types and scales. This often leads to increased processing costs and potential synchronization issues. Second, in centralized network architectures, substantial communication overhead complicates inter-twin knowledge sharing, as transmitting large volumes of multi-modal data from distributed sources to a central server can cause network congestion and latency, undermining real-time responsiveness. In following sections, we first describe network settings using representative examples, and then introduce our UTT workflow designed to address these challenges and enable seamless transformation among NDTs.

\vspace{-0.1in}
\subsection{Network Settings}
In the contemporary urban environment, the widespread presence of cameras has revolutionized how we capture and analyze the dynamic interactions of daily life. These devices, strategically placed along thoroughfares and busy intersections, function not merely as passive observers but as integral components of an increasingly interconnected digital ecosystem. To fully harness the potential of these visual data streams, we consider a comprehensive setting that integrates radio and sensory modalities into the concept of NDTs, where each camera, equipped with an AP, not only capturing visual feeds but also performing signal processing and radio data measurement. This integration can extend to pedestrian devices, as shown in Fig.~\ref{fig: setting}, which communicate with these APs, creating a rich tapestry of multi-modal sensory data. 
The network of cameras, APs, and user equipment lays the groundwork for constructing distributed multi-modal NDTs, designed to comprehensively digitize the entire network environment. 
It paves the way for our UTT framework to support the new transformation capabilities among data modalities, twin models, and corresponding downstream tasks.

\vspace{-.1in}
\subsection{UTT Workflow}

UTT framework inherently encompasses both distributed transformation and multi-modal model transformation principles. As depicted in Fig.~\ref{fig:fed}, it integrates multi-modal data from various sources, including cameras, phone motion sensors, and wireless fine-time measurements, to create a comprehensive modality fusion mechanism that can be decoded back into the original twin formats in a distributed manner. From the multi-modal data transformation perspective, Fig.~\ref{fig:modal} provides a detailed workflow for operating each NDT within UTT framework. Specifically, we assume there are \( N \) local-area twins in the system, each potentially containing up to three task-oriented NDTs, e.g. vision-based user trajectory prediction twin (\( \bm{\theta}_{Traj}^t \)), user position twin (\( \bm{\theta}_{Pos}^t \)), and physical device tracking twin (\( \bm{\theta}_{Track}^t \)). These NDTs, named according to respective tasks, are continuously fed with real-time data from cameras, APs, and user equipment to maintain their twin models over time. 
Since not all task-oriented NDTs may exist, our UTT framework provides an innovative solution for constructing new task-oriented NDTs based on scenario similarities with existing NDTs.

\subsubsection{Local multi-modal data processing}
Initially, the twin models are processed locally. Taking the Trajectory Twin $\bm{\theta}_{Traj}^t$ as an example, the construction process involves a Trajectory Encoder \( \mathcal{E}_{Traj} \), which leverages convolutional layers followed by transformer layers to extract temporal features of the collected multi-modal data and compress them into a fixed-size vector. It starts with several 1-D convolutional layers to capture local dependencies, followed by pooling layers to reduce dimensionality. Subsequently, the transformer layers with multi-head self-attention mechanisms model long-range dependencies, while feed-forward networks are used to refine the feature representation, enhanced by layer normalization and residual connections.
Similarly, other task-oriented twins, such as Position Twin $\bm{\theta}_{Pos}^t$ and Tracking Twin $\bm{\theta}_{Track}^t$, utilize their respective encoders, \( \mathcal{E}_{Pos} \) and \( \mathcal{E}_{Track} \), to extract features, yielding feature vectors \( \mathbf{f}_{Pos} = \mathcal{E}_{Pos}(\mathbf{x}_{Pos}) \) and \( \mathbf{f}_{Track} = \mathcal{E}_{Track}(\mathbf{x}_{Track}) \), where \( \mathbf{x} \) represents the collected raw data by each twin. These feature vectors are designed to have the fixed size of \( d \) throughout the training procedure to ensure consistency.

Next, those initially processed twin models are fused into a modality fusor, where we propose multiple effective model integration methods as shown in Fig.~\ref{fig:modal}. 
Assuming there are $m$ data modalities in the scenario, one common fusion method is \textit{addition} as in Eq.~(\ref{eq:addition}), where the feature vectors are summed element-wise. This method aggregates information from all modalities, prioritizing the feature richness. Addition-based fusion is computationally efficient but often overlooks complex inter-modality relationships.
Another similar operational method is \textit{average}, which takes the element-wise average of the feature vectors, ensuring each data modality contributes equally to model aggregation. This method also helps in balancing the influence of each data resource. 
However, it may dilute significant features unique to individual modalities—for instance, fine-grained spatial patterns captured by LiDAR data or subtle temporal dynamics present in wireless signal traces may be lost when merged into a generalized representation.
\(
    \mathbf{f}_{\text{addition}} = \sum_{i=1}^{m} \mathbf{f}_i, \quad \mathbf{f}_{\text{average}} = \frac{1}{m} \sum_{i=1}^{m} \mathbf{f}_i.
\label{eq:addition}
\)

As an alternative, \textit{concatenation} stacks the representations end-to-end, trying to preserve the individual characteristics of each data modality and allowing the model to learn complex interactions as:
\(
\mathbf{f}_{\text{concatenation}} = [\mathbf{f}_1; \mathbf{f}_2; \mathbf{f}_3; \ldots; \mathbf{f}_m].
\)
Yet, it results in high-dimensional feature vectors that increase computational cost.
In contrast, \textit{multiplication} fusion captures interactions between different modalities by multiplying the representations element-wise, which highlight the joint-source features:
\(
\mathbf{f}_{\text{multiplication}} = \mathbf{f}_1 \times \mathbf{f}_2 \times \mathbf{f}_3 \times \ldots \times \mathbf{f}_m.
\)
Multiplication emphasizes feature interactions, though it can be sensitive to noise and differences in feature scaling.

As a lightweight fusion strategy, the methods in Eq.~(\ref{eq:maxmin}) take the element-wise maximum and minimum, respectively, emphasizing the most and least prominent features from each modality:
\(
\mathbf{f}_{\text{max}} = \max(\mathbf{f}_1, \mathbf{f}_2,  \ldots, \mathbf{f}_m); \mathbf{f}_{\text{min}} = \min(\mathbf{f}_1, \mathbf{f}_2,  \ldots, \mathbf{f}_m).
\label{eq:maxmin}
\)
This max-min approach highlights extreme values, but it might ignores more subtle yet useful information present in the modalities.

Motivated by these pros and cons, we shift the focus to a gating-based fusion method that involves a weighted sum followed by a sigmoid activation function in Eq. (\ref{eq:gating}), allowing the model to dynamically adjust the importance of each data modality, thus offering flexibility in how captured features are combined to form a twin model.
\(
\mathbf{f}_{\text{gating}} = \sigma(W_1 \mathbf{f}_1 + W_2 \mathbf{f}_2 + W_3 \mathbf{f}_3 + \ldots + W_m \mathbf{f}_m).
\label{eq:gating}
\)
This gating-based fusion dynamically adapts the weighting of modalities, making it the best method due to its superior flexibility and ability to suppress noise while emphasizing the most informative features.
Similarly, an attention mechanism can be exploited to compute a weighted sum, where the weights are determined by the relevance of each feature vector, providing an adaptive way to integrate multi-modal data, emphasizing the most informative features as in:
\(
\mathbf{f}_{\text{attention}} = \sum_{m} \bm{\theta}_m \mathbf{f}_m, \quad \bm{\theta}_m = \frac{\exp(e_m)}{\sum_{j} \exp(e_j)}.
\)
Such an attention-based fusion offers context-aware weighting, although its performance is highly dependent on the quality of the learned attention scores.

All above fusion variations from Eq. (1) to Eq. (6) have the potential to ensure the effective integration of diverse sensory data, enhancing the richness of the combined model representation. We compare and evaluate the performance of these fusion techniques within our UTT framework in Sec. 5.
Particularly, the results show that the gating-based method offers the most effective solution for adaptive fusion in our considered context, while other methods serve as useful alternatives or complementary approaches.

After this model fusion stage, the abstracted knowledge vectors from the fusor are decoded back into the original formats using corresponding decoders. Each decoding module consists of transformer layers followed by deconvolutional layers, reconstructing new NDTs to match the input data modalities. This sophisticated architecture facilitates the integration and synthesis of task-oriented NDTs, enabling the network system to capture complex interactions across different sensory data. 

\subsubsection{Distributed model-level mapping}
Based on the local data fusion steps, our UTT framework employs distributed mapping to perform further model aggregations across different NDTs. Specifically, each deployed device collects multi-modal data from various locations and synchronizes with its corresponding NDT, ensuring that the mapping process benefits from diverse data sources and environmental areas. The local-area NDTs first send their twin models to a central server, where model aggregations are performed to construct a global task twin, as shown in Fig.~\ref{fig:fed}. This updated global task twin is then synchronized back to each local-area NDT for the next round of mapping and/or update. 
Such a distributed learning architecture also enhances data privacy by keeping raw data modalities localized and sharing only model parameters, which is crucial for maintaining data sensitivity in a practical setting. 
The overall distributed multi-modal mapping process is shown in Algorithm~\ref{alg:mapping}, which mainly consists of six stages:

\begin{algorithm}
\caption{UTT Enabled Mapping Process}
\label{alg:mapping}
\begin{algorithmic}[1]

\Require 
\Statex The number of local-area NDT $n$, number of global training rounds $T$, local data sets $D_{i,m}^{t}$, initial global model  $\bm{\theta}^{0}$, encoders $\{\mathcal{E}_m\}_{m \in M}$, decoders $\{\mathcal{D}_m\}_{m \in M}$

\Ensure 
\Statex Optimal global twin model $\bm{\theta}^{T}$

\For{$t = 0,\ldots,T-1$}

    \For{each local-area NDT $i = 1,\ldots,n$}
        \For{each modality $m$}
            \State $\mathbf{x}_{i,m}^{t} \gets D_{i,m}^{t}$ 
            \State $\mathbf{f}_{i,m}^{t} \gets \mathcal{E}_m(\mathbf{x}_{i,m}^{t})$ 
        \EndFor

        \State $\mathbf{f}_{i,\text{fusion}}^{t} \gets f_{\text{fusion}}\bigl(\{\mathbf{f}_{i,m}^{t}\}_{m \in M}\bigr)$
    \EndFor

    \For{each local-area NDT $i = 1,\ldots,n$}
        \State $\bm{\theta}_i^{t} \gets \bm{\theta}^{t}$ 
        
        \For{each modality $m \in M$}
            \State $\mathbf{\hat{f}}_{i,m}^{t} \gets \mathcal{D}_m(\mathbf{f}_{i,\text{fusion}}^{t})$ 
        \EndFor
        
        \State $\mathcal{L}_i^{t} \gets \sum_{m \in M} \mathcal{L}\bigl(\mathbf{\hat{f}}_{i,m}^{t}, \mathbf{f}_{i,m}^{t}\bigr)$
        \State $\bm{\theta}_i^{t} \gets \text{train}\bigl(\bm{\theta}_i^{t}, \mathcal{L}_i^{t}\bigr)$ 
        
    \EndFor

    \For{each local-area NDT $i = 1,\ldots,n$}
        \State upload $\bm{\theta}_i^{t}$ to central server
    \EndFor

    \State $\bm{\theta}^{t+1} \gets \frac{1}{n}\sum_{i=1}^{n}\bm{\theta}_i^{t}$

\EndFor

\State \Return $\bm{\theta}^{T}$

\end{algorithmic}
\end{algorithm}

\begin{list}{\labelitemi}{\leftmargin=1em \itemindent=-0.08em \itemsep=.2em}
    \item \textbf{Stage I (Local-area Data Processing): }
    Each local-area NDT \( i \) gathers data across existing task-oriented NDTs. These data are transformed into the feature vectors \( \mathbf{f}_i \) using specialized feature extraction functions \( \phi_i \). This stage captures essential information from each data modality in a structured format (Lines 3-6). 

    \item \textbf{Stage II (Feature Vector Fusion): }
    The extracted feature vectors from different modalities \( m \in M \) are combined into a unified representation \( \mathbf{f}_{\text{fusion}} \). 
    A common fusion method is concatenation, merging vectors into a single comprehensive vector.
    This consolidated feature vector encapsulates information from all data modalities for subsequent model analysis (Line 7).

    \item \textbf{Stage III (Initial Local Model Training): }
    Each twin utilizes the fused feature vector \( \mathbf{f}_{\text{fusion}} \) to train a local model \( \bm{\theta}_i \). This model is initialized with the current global model parameters \( \bm{\theta}^t \). The training process involves minimizing a loss function \( \mathcal{L} \) that quantifies the error between the model’s predictions and the actual measurements based on the local data (Lines 9-16). 

    \item \textbf{Stage IV (Local Model Upload): }
    After the local training phase, each local-area NDT transmits its twin model parameters \( \bm{\theta}_i^{t} \) to the global twin (Lines 17-19).

    \item \textbf{Stage V (Global Model Synchronization): }
    The global twin aggregates the received local-area models to update the global twin model parameters. 
    The global twin model is computed by averaging the parameters of all local-area NDTs. 
    Note that this aggregation ensures that the global twin model incorporates a collective insight from participating local-area twins (Line 20).

    \item \textbf{Stage VI (Iterative Refinement): }
    The updated global model \( \bm{\theta}^{t+1} \) is redistributed to all local-area twins for further tuning rounds. This iterative process is repeated until the global twin achieves satisfactory performance in twinning accuracy and/or downstream tasks. Each iteration involves repeating Stages III to V, continuously refining the global twin through successive cycles of local training and global aggregation (back to Line 1).
    
\end{list}

\vspace{-0.1in}

\subsection{Twin-to-Twin Transformation}

Building upon the above workflow, 
UTT framework is tamed to construct new NDTs from existing twin models by efficiently transferring, merging, and splitting NDT models across various data modalities. 
The main twinning functions achieved by UTT are categorized in Fig.~\ref{fig:multi_modal_learning}, enabling both data modality and downstream task transitions within NDTs.

\subsubsection{Twin-to-Twin Transfer}
Within UTT framework, the transfer of NDT is a critical functionality that enables the extension of existing NDT models to new domains or modalities. For instance, if we have an existing Position Twin, UTT can facilitate the transfer of this twin model to a new twin model, such as Trajectory Twin or Tracking Twin, using the captured and abstracted data features, as demonstrated in Fig.~\ref{fig:multi_modal_learning} (a) and (b). This transfer capability is crucial for expanding the utility of a single, well-developed NDT for different network tasks.

\subsubsection{Multi-Twin Merging}
The merging capability of UTT addresses the scenarios where multiple NDTs, such as the Trajectory Twin and the Tracking Twin, have already existed in the network system. By integrating these twins, our UTT framework can generate a new NDT such as the Position Twin by associating with the fused data features from those existing twins, as demonstrated in Fig.~\ref{fig:multi_modal_learning} (c). This merging process is vital for synthesizing data from disparate sources, leading to a more unified and powerful representation of the network. In essence, such an integrative capability enables the creation of complex, multi-faceted NDTs that reflect the interconnected nature of physical environment, thereby enhancing new predictive analytics and tasks.

\subsubsection{Single-Twin Splitting}
Conversely, UTT framework also provides the intra-twin functionality to split an existing NDT into several task-oriented twins, as demonstrated in Fig.~\ref{fig:multi_modal_learning}(d). This capability is particularly useful for isolating specific data modalities or functionalities for isolated operations. Splitting one task-oriented NDT, such as a Position Twin, involves reforming its underlying data pattern, either from one modality to another or from one local-area model to another, through the joint local data processing and distributed mapping as in Sec. III-B. This operation can be replicated to split various types of NDTs, making it a versatile tool for generalizing network functions from a single powerful twin.

To implement UTT framework with all aforementioned twin-to-twin transformations, we consider two model training approaches: 1) train a unified UTT model that generalizes twin transfer, merging, and splitting operations, or 2) train three function-specific models, one for each operation. 
The main difference lies in the loss function design in the Stage III of UTT workflow. Specifically, the loss function of the unified UTT model can be expressed as 
\(
\mathcal{L}_{\text{UTT}} = \mathcal{L}_{T} + \mathcal{L}_{M} + \mathcal{L}_{S},
\)
where \(\mathcal{L}_{T}\), \(\mathcal{L}_{M}\), and \(\mathcal{L}_{S}\) represent the individual loss functions for the transfer, merging, and splitting operations, respectively. For the function-specific models, each twin model has its own loss function designed dedicated for that operation. For instance, the loss function for the transfer function model is 
\(
\mathcal{L}_{T} = \text{Loss}(\bm{\theta}_{T}(\mathbf{f}_{\text{fusion}}), y_{T}).
\)
This approach allows for tailored optimization of each specific twin operation, enhancing the twinning performance in scenarios where existing NDTs have significantly different characteristics.
We will evaluate the performance of these two modeling approaches in Sec. 5.

\section{Theoretical Analysis on Twin Transformation}

In this section, we provide a theoretical guarantee for our UTT framework by establishing the convergence bounds of the proposed multi-modal, distributed mapping process. These results demonstrate that the transformation method can consistently construct new twins within a bounded error range, ensuring stability and reliability across varying input modalities and network conditions.



\begin{assumption}[Lipschitz Gradients]
Each $F_i$ is $L$-smooth, i.e., $\|\nabla F_i(x) - \nabla F_i(y)\| \leq L \|x - y\|$ for all $x,y$. 
\label{asp:lipschitz}
\end{assumption}

\begin{assumption}[Bounded Gradients]
The coordinate-wise gradients satisfy $|[\nabla f_i(x,z)]_j| \leq G$ for all $j$.
\label{asp:bounded-grad}
\end{assumption}

\begin{assumption}[Bounded Variance]
For each coordinate $j$, the local variance satisfies $\mathbb{E}[\|[\nabla f_i(x,z)]_j - [\nabla F_i(x)]_j\|^2] = \gamma_{l,j}^2$. Furthermore, the global variance is bounded as $\frac{1}{n}\sum_{i=1}^n \|[\nabla F_i(x)]_j - [\nabla f(x)]_j\|^2 \leq \gamma_{g,j}^2$.
\label{asp:variance}
\end{assumption}



The overall analysis is based on three standard assumptions in non-convex optimization problems~\cite{50448, 8664630, NEURIPS2018_90365351}.
%
At iteration $t$, let $\bm{\theta}_t \in \mathbb{R}^d$ be the current global parameter. For each modality $m \in \{1, \ldots, M\}$, the twin model update follows
\(
\omega_t^{(m)} = \epsilon \omega_{t-1}^{(m)} + (1 - \epsilon) \left(\Delta_t^{(m)}\right)^2,
\)
where $f^{(m)}$ is the loss function associated with data modality $m$, the overall objective is $f(x) = \sum_{m=1}^M f^{(m)}(x)$, and local twin update is $\Delta_t$. We consider a positive constant $\mu > 0$ and a learning rate $\eta > 0$. The updates from different modalities are integrated using a gating fusion mechanism, as defined in Eq.~(\ref{eq:gating}). This approach is demonstrated to be the most effective mechanism in our subsequent experiments (Sec. 5.1). As such, the global parameter updates follows
\(
\bm{\theta}_{t+1} = \bm{\theta}_t + \eta \sum_{m=1}^M \alpha_m \frac{\Delta_t^{(m)}}{\sqrt{\omega_t^{(m)}} + \mu},
\)
where $\alpha_m \geq 0$ and $\sum_{m=1}^M \alpha_m = 1$.


\begin{theorem}
We define $\beta$ as the number of local training rounds, $N$ as the number of global training rounds, $\eta_l$ as the twin model-level learning rate, and $\eta$ as the global NDT update rate. Suppose the assumptions above hold and let $\gamma^2 = \sum_{j=1}^d (\gamma_{l,j}^2 + 6 \beta \gamma_{g,j}^2)$. If $\eta_l$ satisfies that
\[
\eta_l \leq \frac{1}{16\beta} \min\left\{\left(\frac{\mu}{120GL^2}\right)^{1/3},\; \frac{\mu}{4G+2\eta L}\right\},
\]
then under the gating-based scheme in UTT framework, the average rate of decrease in the norm of the gradient satisfies
\begin{align*}
    &\min_{0 \leq t \leq N-1} \mathbb{E}\|\nabla f(\bm{\theta}_t)\|^2 \\
    &\quad\;=\; O\Biggl(\frac{\sqrt{\epsilon \eta_l \beta G} + \mu}{\eta_l \beta N} \Biggl( \frac{f(x_0) - f(x^*)}{\eta} + \frac{5\eta_l^3 \beta^2 L^2 N}{2\mu}\gamma^2 \\
    &\quad\;+ \left(G + \frac{\eta L}{2}\right) \Biggl[ \frac{4\eta_l^2 \beta N}{m \mu^2} \sum_{j=1}^d \gamma_{l,j}^2 + \frac{20\eta_l^4 \beta^3 L^2 N}{\mu^2} \gamma^2 \Biggr] \Biggr) \Biggr).
\end{align*}
\end{theorem}

In other words, under suitable parameter choices, 
the algorithm can drive the expected gradient norm toward zero at a rate captured by a complexity bound. Thus, the proposed multi-modal gated fusion method does not affect the convergence bound theoretically.

\vspace{-0.1in}
\begin{proof}
Recall that $f$ is $L$-smooth. Substituting the update rule with
\(
A = \sum_{m=1}^M \alpha_m \frac{\Delta_t^{(m)}}{\sqrt{\omega_t^{(m)}} + \mu}
\)
into $L$-smoothness inequality, we have:
\vspace{-0.4cm}
\begin{align*}
    f(\bm{\theta}_{t+1}) \leq f(\bm{\theta}_t) \;&+\; \eta \langle \nabla f(\bm{\theta}_t), A \rangle 
    + \frac{\eta^2 L}{2} \sum_{i=1}^{d} A_i^2.
\end{align*}
\noindent Taking expectations given $\bm{\theta}_t$, and introducing \\
\(B = \sum_{m=1}^M \alpha_m \frac{\Delta_t^{(m)}}{\sqrt{\epsilon \omega_{t-1}^{(m)}} + \mu},\)
\vspace{-0.2cm}
we obtain:
\begin{align*}
    \mathbb{E}_t[f(\bm{\theta}_{t+1})] \;&\leq\; f(\bm{\theta}_t) \;+
    \eta \Bigg\langle \nabla f(\bm{\theta}_t), \mathbb{E}_t\left[A \;-B \right]\Bigg\rangle \\
    &+\; \eta \Bigg\langle \nabla f(\bm{\theta}_t), \mathbb{E}_t\left[B\right]\Bigg\rangle +\; \frac{\eta^2 L}{2}\sum_{j=1}^d \mathbb{E}_t\left[A^2\right].
\end{align*}

By defining
\(
C = \sum_{j=1}^d \frac{[\nabla f(\bm{\theta}_t)]_j^2}{\sqrt{\epsilon \omega_{t-1,j}} + \mu},
\)
the quantity $C$ acts as a measure of the weighted gradient norm, reflecting the influence of the adaptive denominators.
First, the second term in the equation gives:
\begin{align*}
    &\Bigg\langle \nabla f(\bm{\theta}_t), \mathbb{E}_t\left[A \;-B \right]\Bigg\rangle 
    = \mathbb{E}_t \sum_{j=1}^d [\nabla f(\bm{\theta}_t)]_j \Bigg[A-B\Bigg] \\
    &\leq (1-\epsilon)\mathbb{E}_t \sum_{j=1}^d |\nabla f(\bm{\theta}_t)|_j \sum_{m=1}^M \alpha_m \\
    &\frac{(\Delta_{t,j}^{(m)})^2}{(\sqrt{\omega_{t,j}^{(m)}}+\mu)\bigl(\sqrt{\epsilon \omega_{t-1,j}^{(m)}}+\mu\bigr)\bigl(\sqrt{\epsilon \omega_{t-1,j}^{(m)}}+\sqrt{\omega_{t,j}^{(m)}}\bigr)}. 
\end{align*}
This shows that the difference $(A-B)$ is controlled by factors involving the bounded gradients, the adaptive terms, and $\sqrt{1-\epsilon}$.
Next, we consider:
\vspace{-0.1cm}
\begin{align*}
     \Bigg\langle \nabla f(\bm{\theta}_t), \mathbb{E}_t\left[B\right]\Bigg\rangle 
    =\; &-\eta_l \beta C + \Bigg\langle \frac{\nabla f(\bm{\theta}_t)}{\sqrt{\epsilon \omega_{t-1}} + \mu}, \\
    &\mathbb{E}_t\left[\sum_{m=1}^{M}\alpha_m \Delta_t^{(m)} + \eta_l \beta\nabla f(\bm{\theta}_t)\right]\Bigg\rangle.
\end{align*}
Next, we define
\(
D = \sum_{m=1}^M \frac{\alpha_m}{\sqrt{\epsilon \omega_{t-1,j}^{(m)}} + \mu},
\)
and 
\(
\gamma^2 = \sum_{j=1}^d (\gamma_{l,j}^2 + 6\beta\gamma_{g,j}^2).
\)
Using the bounded variance and smoothness assumptions and inserting the above inequality back into the expression, and combining these results and summing from $t=0$ to $N-1$ yields:
\begin{align*}
    &\mathbb{E}_t[f(\bm{\theta}_t)] \;\leq\; f(x_0) \;
    - \frac{\eta \eta_l \beta}{8} \sum_{t=0}^{N-1}\sum_{j=1}^d [\nabla f(\bm{\theta}_t)]_j^2 D \\
    &+ \frac{5\eta \eta_l^3 \beta^2 L^2 N}{2\mu} \mathbb{E}\left[\gamma^2\right] + \left(\eta \sqrt{1-\epsilon}G + \frac{\eta^2 L}{2}\right) \cdot \\
    &\Bigg[\frac{4\eta_l^2 \beta N}{m \mu^2}\sum_{j=1}^d \gamma_{l,j}^2 
    + \frac{20\eta_l^4 \beta^4 L^2 N}{\mu^2}\mathbb{E}\left(\gamma^2\right)\Bigg].
\end{align*}
The final convergence bound can be derived by integrating this inequality with previous bounds through reformulation as:
\begin{align*}
    \sum_{t=0}^{N-1} C
    \;&\geq\; \frac{N}{\sqrt{\epsilon \eta_l \beta G} + \mu} \min_{0 \leq t < N}\|\nabla f(\bm{\theta}_t)\|^2.
\end{align*}
\end{proof}

\vspace{-0.1in}
\section{Experimentation and Evaluations}
\label{sec:exp}

\subsection{Twin-to-Twin Operations}

\begin{figure*}[ht]
    \centering
    \begin{subfigure}{0.3\textwidth}
        \centering
        \includegraphics[width=\linewidth]{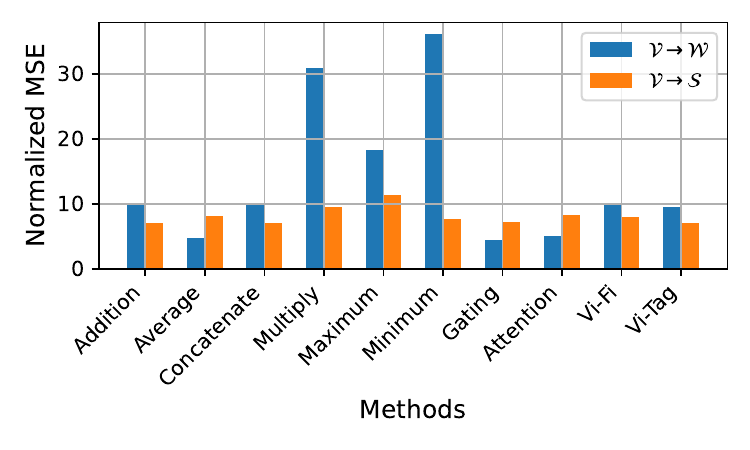} 
        \vspace{-0.5cm}
        \caption{Transfer from Trajectory Twin.}
        
        \label{fig:first}
    \end{subfigure}
    \begin{subfigure}{0.3\textwidth}
        \centering
        \includegraphics[width=\linewidth]{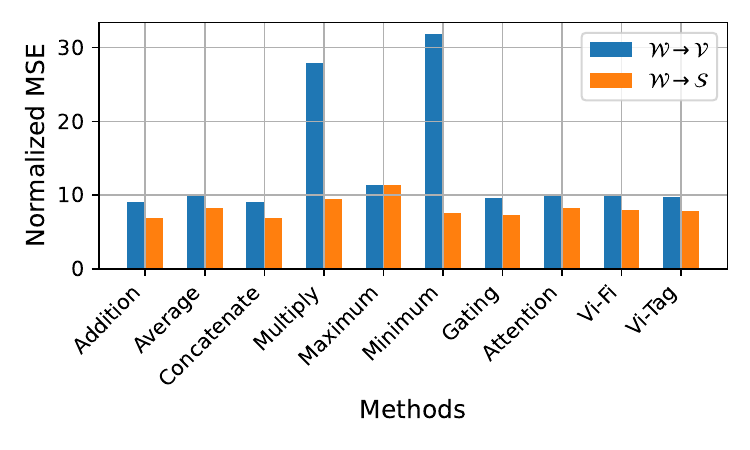} 
        \vspace{-0.5cm}
        \caption{Transfer from Position Twin.}
        \label{fig:second}
    \end{subfigure}
    \begin{subfigure}{0.3\textwidth}
        \centering
        \includegraphics[width=\linewidth]{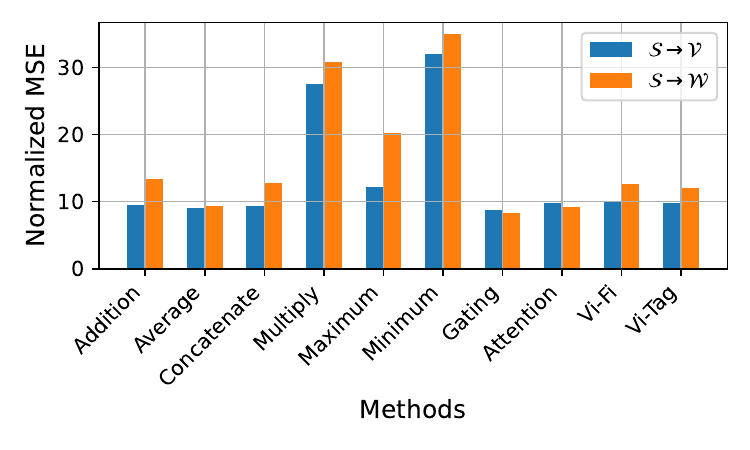} 
        \vspace{-0.5cm}
        \caption{Transfer from Tracking Twin.}
        \label{fig:third}
    \end{subfigure}

    \begin{subfigure}{0.3\textwidth}
        \centering
        \includegraphics[width=\linewidth]{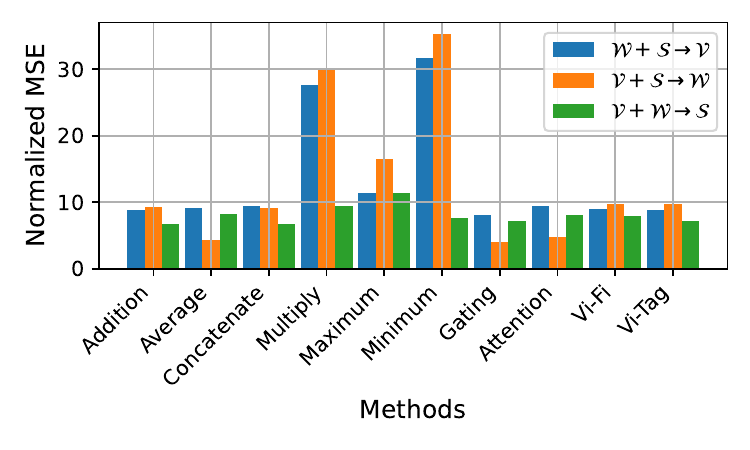} 
        \vspace{-0.5cm}
        \caption{Multi-twin Merging.}
        \label{fig:fourth}
    \end{subfigure}
    \begin{subfigure}{0.3\textwidth}
        \centering
        \includegraphics[width=\linewidth]{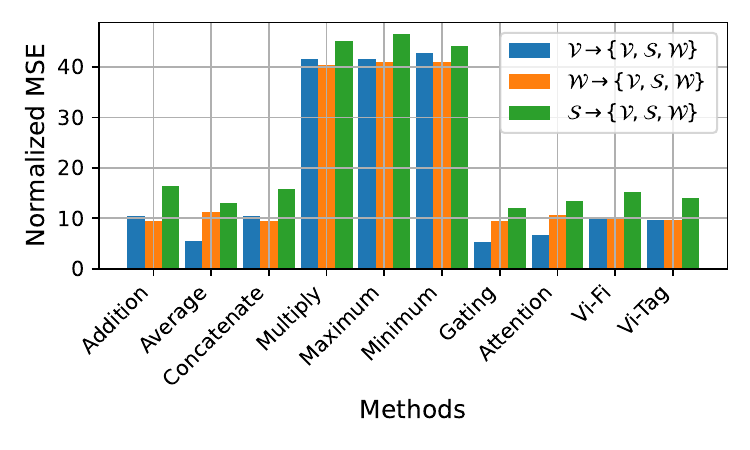} 
        \vspace{-0.5cm}
        \caption{Single-twin Splitting.}
        \label{fig:fifth}
    \end{subfigure}
\vspace{-.1in}
    \caption{Normalized MSE of twin-to-twin transformation of UTT framework.}
    \label{fig:num_results}
    \vspace{-.2in}
\end{figure*}

\subsubsection{Experiment Setup}

The evaluations use a multi-modal dataset, as introduced in~\cite{9918171}, to assess the efficacy of twin transitions with different data modalities using our UTT framework. This extensive dataset encompasses vision, wireless, and smartphone motion sensor data from multiple fixed devices and mobile users in both indoor and outdoor settings. Specifically, the vision data modality includes RGB-D video captured by a mounted camera as in Fig.~\ref{fig: setting}, while the radio data modality features the signals from users' smartphones, including wireless Fine Time Measurement (FTM) and Inertial Measurement Unit (IMU) sensory readings. The real-world data collection setup employed a Stereolabs ZED2 (RGB-D) camera, recording video at 10 fps and capturing depth information ranging from 0.2m to 20m. Smartphones are configured to exchange FTM messages at a frequency of 3Hz with a Google Nest Wi-Fi AP positioned next to the camera. 
The collected vision, radio, and sensory data contribute to three NDTs that support the downstream predictive tasks of user trajectory prediction, positioning, and device tracking.

\subsubsection{Baseline Methods}
As this is the first work to explore the twin-to-twin transformation process among NDTs, there are no directly comparable baselines. To enable a meaningful evaluation, we re-implement and assess two model reconstruction methods that approach the problem from the perspective of data-level multi-modal learning and transformation.
The first method, Vi-Fi, as detailed in~\cite{9826015}, utilizes a model-driven bipartite matching algorithm to align device trajectories estimated from camera views with those obtained from wireless phone sensor measurements. The second method, Vi-Tag, described in~\cite{9918171}, introduces a multi-modal LSTM-based autoencoder designed to learn a representation between different modalities during model training. For the purposes of our twin-to-twin transformation, both baseline methods have been reproduced within our distributed mapping framework to ensure consistency and comparability.

\subsubsection{Performance of UTT Framework}

\begin{table}[b]
\centering
\footnotesize
\setlength\extrarowheight{2.5pt}
\caption{ 
Accuracy and overhead comparison between direct mappings and our twin transformations. 
}
\begin{tabular}{>{\raggedright}p{3.3cm}|*{2}{>{\centering\arraybackslash}p{2.2cm}}}
\toprule
 & Time (Sec) & Accuracy (NMSE)\\
\midrule
Direct mappings~\cite{zhang2024mapping}       & 6,843 & 15.79 \\
UTT twin splitting           & 4,002 & 12.01 \\ 
Centralized twin splitting   & 4,571 & 12.61 \\
UTT twin merging             & 4,355 & 8.08 \\ 
Centralized twin merging     & 4,941 & 8.24 \\

\bottomrule
\end{tabular}
\label{tab:cost}
\vspace{-0.1in}
\end{table}


Fig.~\ref{fig:num_results} shows the normalized Mean Squared Error (MSE) between twin-generated data and ground truth for various proposed modality fusors and across different types of twin-to-twin transformations. The ground-truth dataset including all sensed human activities is collected from~\cite{9918171}. Figs.~\ref{fig:first} to~\ref{fig:third} correspond to the NDT transferring, while Figs.~\ref{fig:fourth} and~\ref{fig:fifth} show the performance of the NDT merging and splitting operations. For simplicity, we use the modal abbreviations $\mathcal{V}$ (Visual), $\mathcal{W}$ (Wireless), and $\mathcal{S}$ (Sensory) to correspond to Trajectory, Position, and Tracking twins, respectively. Arrows are used to denote the transformation directions.

First, in Figs.~\ref{fig:first},~\ref{fig:second}, and~\ref{fig:third}, we observe that the proposed gating fusor shows significantly lower MSE values for all transformation operations compared to other fusion methods. For example, in Fig.~\ref{fig:first}, the normalized MSE for the gating fusor in $\mathcal{V} \rightarrow \mathcal{W}$ is 4.49, while in $\mathcal{V} \rightarrow \mathcal{S}$ it is 7.18. Similarly, in Fig.~\ref{fig:second}, the gating fusor outperforms other methods for $\mathcal{W} \rightarrow \mathcal{V}$ and $\mathcal{W} \rightarrow \mathcal{S}$ transformations with normalized MSE values of 9.65 and 7.58, respectively. Fig.~\ref{fig:third} also shows the similar comparison trends when performing $\mathcal{S} \rightarrow \mathcal{V}$ and $\mathcal{S} \rightarrow \mathcal{W}$. These results demonstrate the effectiveness of the gating fusion method adopted in our UTT framework for the twin-to-twin transfer, which ensures the model accuracy across different NDTs. This is because the gating fusor dynamically adjusts the weights assigned to each modality, thereby capturing the most relevant features for the transformation process.
Second, regarding the multi-twin merging and single-twin splitting operations as shown in Figs.~\ref{fig:fourth} and~\ref{fig:fifth}, it is observed that the gating fusor in UTT framework maintains its superiority over other methods as well as baselines. Particularly, for the merging operation of $\mathcal{W}+\mathbf{S} \rightarrow \mathcal{V}$, $\mathcal{V}+\mathbf{S} \rightarrow \mathcal{W}$, and $\mathcal{V}+\mathbf{W} \rightarrow \mathcal{S}$, our gating method consistently shows lower MSE values, such as 8.04, 4.00, and 7.15, respectively. This is in contrast to methods like Multiply and Maximum, which always result in higher error rates, perhaps due to their inability to effectively balance the contributions from different data modalities, leading to either overemphasis on noise or under-utilization of useful information. The gating fusor leverages an adaptive mechanism that fine-tunes the fusion process, enhancing the overall quality of the merged data.
Fig.~\ref{fig:fifth} further confirms the feasibility of the twin splitting functionality, where $\mathcal{V} \rightarrow \{\mathcal{V},\mathcal{S},\mathcal{W}\}$, $\mathcal{W} \rightarrow \{\mathcal{V},\mathcal{S},\mathcal{W}\}$, and $\mathcal{S} \rightarrow \{\mathcal{V},\mathcal{S},\mathcal{W}\}$ show fairly lower MSE values. 
In addition, Tab.~\ref{tab:cost} illustrates the communication and computational costs. “Direct mappings” refer to the baseline in which each twin is constructed solely from its native data modality, with no exposure to the other twin models. For splitting and merging, we evaluate the worst-case configurations, that $S\rightarrow \{V,S,W\}$ for splitting and $W+S\rightarrow V$ for merging. In a centralized setting, all raw data is uploaded to the server for direct mapping, replacing the multi-modal, federated workflow. “Time consumption (s)” aggregates both computation (data preprocessing and model twinning) and communication (model and data transfer) overheads. It is observed that our proposed UTT framework does not add extra complexity to the twinning process. This is because the direct mapping excludes cross-modal fusion, each twin must rely on its own modality, and the absence of complementary information increases twinning error. The procedure is nevertheless faster, as one mapped twin model can be reused to instantiate the remaining twins, eliminating the overhead of running three separate mapping pipelines (e.g. for splitting cases). Besides, the federated mapping design reduces the communication overhead significantly, since raw data is not always transmitted at all.

\begin{figure}[t]
    \centering
    \begin{subfigure}[b]{0.23\textwidth}
        \centering
        \includegraphics[width=\textwidth]{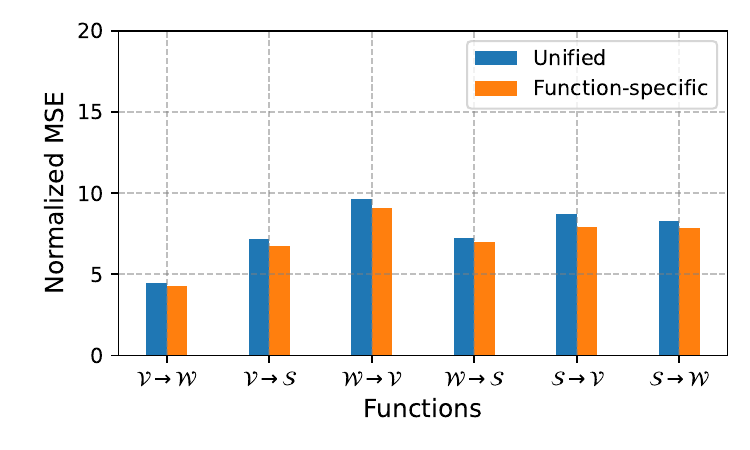}
        \hfill
        \caption{Impact of different models in transfer.}
        \label{fig: general_1}
    \end{subfigure}
    \hfill
    \begin{subfigure}[b]{0.23\textwidth}
        \centering
        \includegraphics[width=\textwidth]{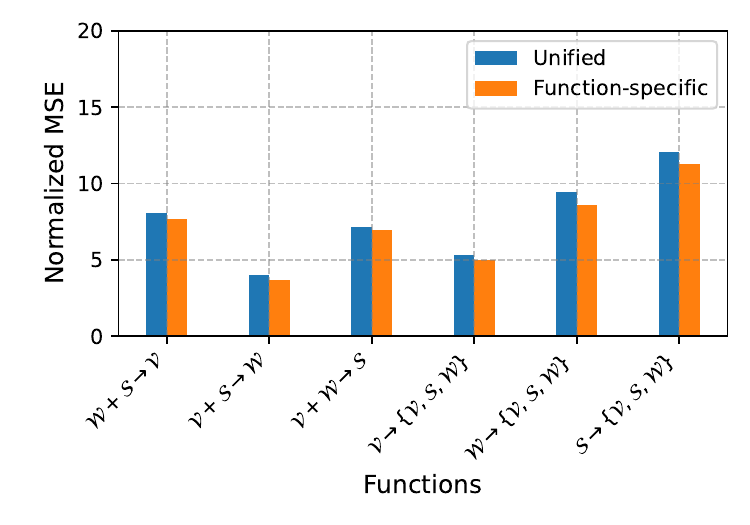}
        \caption{Impact of different models in merging and splitting.}
        \label{fig: general_2}
    \end{subfigure}
    \caption{Impact of different UTT models in twin operations.}
    \label{fig:general_combined}
    \vspace{-.15in}
\end{figure}


\subsubsection{Unified vs. Function-specific UTT Model(s)}

Recall our UTT framework, as depicted in Fig.~\ref{fig:modal}, there are two approaches to train the transformation model for twin transfer, merging, and splitting operations: 1) train a unified model for all operations, or 2) train three function-specific models, one for each operation. For the unified model, the training loss function is designed to construct a general-purpose model for all twin operations, with the global loss function being the sum of the loss functions for each specific operation. In contrast, function-specific modeling involves training a dedicated model for each specific operation, such as transfer, merging, or splitting. Figs.~\ref{fig: general_1} and~\ref{fig: general_2} explore the performance difference between the unified and function-specific modeling approaches. As expected, the results show that the function-specific models generally yield better performance, as indicated by lower normalized MSE values, due to the model personalization for each specific function. For instance, in Fig.~\ref{fig: general_1}, the normalized MSE for the function-specific model in $\mathcal{V} \rightarrow \mathcal{W}$ is 4.30 compared to 4.49 for the unified transformation model. However, such an approach is limited to specific operations like twin-to-twin transfer and cannot be generalized to other operations. This highlights the need to balance model accuracy and implementation efficiency, with the unified model being more efficient for broader twin transformations despite a slight degradation in twinning accuracy. 


\begin{figure}[h]
	\centering
	\includegraphics[scale = 0.5]{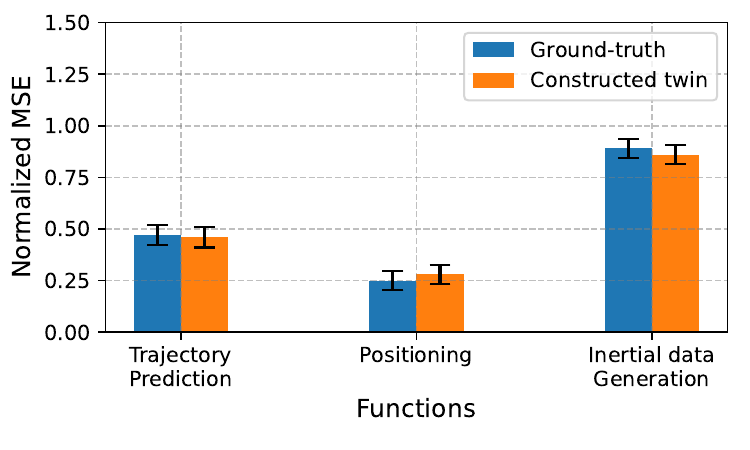}
  \vspace{-.1in}
	\caption{Normalized MSE for various downstream tasks within the constructed NDTs.}
	\label{fig: tasks}
\end{figure}

\begin{figure*}[t]
	\centering
	\includegraphics[scale = 0.7]{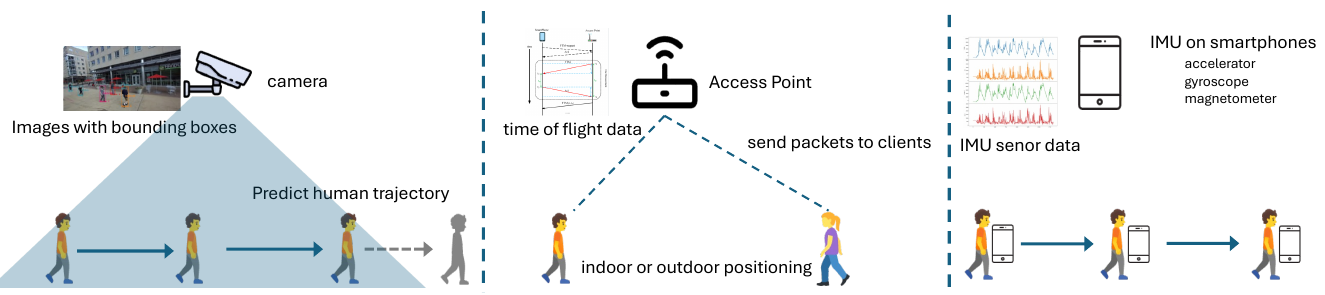}
	\caption{Experiments of NDT-assisted downstream tasks.}
	\label{fig: task_setting}
	\vspace{-.2in}
\end{figure*}



\vspace{-0.1in}
\subsection{NDT-assisted Downstream Tasks}

Lastly, we apply our constructed NDTs from UTT framework to real-world applications, including human trajectory prediction, indoor and outdoor positioning, and inertial data generation. 
The scenario configurations are demonstrated in Fig.~\ref{fig: task_setting}. 
Specifically, human trajectory prediction involves using mapped context and visual data from NDT to predict future movements of a person based on their past trajectory. This is achieved by analyzing sequences of virtual image data in the twin to extract motion patterns and predict likely path a person will take. 
Second, indoor and outdoor positioning is performed using radio FTM from the corresponding NDT-assisted data measurements, which provides precise object location information by measuring the time it takes for a wireless signal to travel between two virtual entities in NDT.  
In practice, data from FTM can be integrated with other sensory data to enhance positioning accuracy and provide more robust localization services, which demonstrates the potential of our multi-twin merging technique to integrate an FTM-based twin and a sensory twin.
The third evaluation task is inertial data generation for physical device tracking, which utilizes data from IMU sensors within the NDT coverage areas to track the human pose and movements. 
Combining IMU data with sensory inputs like cameras allows the merged NDT to enable advanced functionalities such as comprehensive device tracking, gesture recognition, and activity monitoring. 

Fig.~\ref{fig: tasks} shows the comparison of aforementioned task performance from our constructed NDTs to the ground-truth measurements. It is noted that the performance gaps between the ground-truth measurements and the generated results from the NDTs are quite small across all three tasks. For the human trajectory prediction, the normalized MSE for real and NDT measurements are nearly identical, which are 0.474 and 0.462 respectively, indicating that our constructed NDT can accurately replicate the patterns observed in a real-world environment. Similarly, for the positioning and sensory data generation tasks, the normalized performance scores closely match those of the ground-truth measurements, demonstrating reliability of constructed twins through UTT framework.

\section{Conclusion}

In this paper, we introduce a pioneering unified twin transformation framework that advances the efficient construction of NDTs. 
This enables the transfer, merging, and splitting of existing NDTs to create task-oriented NDTs by integrating knowledge from multiple modalities, enhancing accuracy, scalability, and resource efficiency without constructing twins from scratch.
Comprehensive evaluations of twin-to-twin transformation accuracy and downstream DT-assisted applications, such as trajectory reconstruction, positioning, and sensory data generation, underscore the practical utility and robustness of our framework in real-world scenarios. 

\section{Acknowledgment}

This research was supported by the National Science Foundation through Award CNS--2440756, CNS--2312138 and CNS--2312139.

\bibliographystyle{IEEEtran}
\bibliography{refs}

\end{document}